\documentclass[aps,prl,twocolumn,showpacs,superscriptaddress]{revtex4-1}

\usepackage[center]{subfigure}
\usepackage[utf8x]{inputenc}
\usepackage{fullpage}
\usepackage{sidecap}
\usepackage[pdftex]{graphicx}
\usepackage{amsmath}
\usepackage{units}

\begin{document}

\title{Identification of drug resistance mutations in HIV from constraints on natural evolution}
\author{Thomas C Butler}
\thanks{These authors contributed equally to this work}
\affiliation{Department of Physics, Massachusetts Institute of Technology, Cambridge, MA, USA}
\affiliation{Department of Chemical Engineering, Massachusetts Institute of Technology, Cambridge, MA, USA}
\author{John P Barton}
\thanks{These authors contributed equally to this work}
\affiliation{Department of Physics, Massachusetts Institute of Technology, Cambridge, MA, USA}
\affiliation{Department of Chemical Engineering, Massachusetts Institute of Technology, Cambridge, MA, USA}
\affiliation{Ragon Institute of Massachusetts General Hospital, Massachusetts Institute of Technology and Harvard University, Boston, MA, USA}
\author{Mehran Kardar}
\email[]{kardar@mit.edu}
\affiliation{Department of Physics, Massachusetts Institute of Technology, Cambridge, MA, USA}
\author{Arup K Chakraborty}
\email[]{arupc@mit.edu}
\affiliation{Department of Physics, Massachusetts Institute of Technology, Cambridge, MA, USA}
\affiliation{Department of Chemical Engineering, Massachusetts Institute of Technology, Cambridge, MA, USA}
\affiliation{Ragon Institute of Massachusetts General Hospital, Massachusetts Institute of Technology and Harvard University, Boston, MA, USA}
\affiliation{Departments of Chemistry and Biological Engineering, Institute for Medical Engineering and Science, Massachusetts Institute of Technology, Boston, MA, USA}

\newpage

\begin{abstract}

Human immunodeficiency virus (HIV) evolves with extraordinary rapidity.  However, its evolution is constrained 
by interactions between mutations in its fitness landscape. 
Here we show that an Ising model describing these interactions, inferred from 
sequence data obtained prior to the use of antiretroviral drugs, can be used to identify 
clinically significant sites of resistance mutations. 
Successful predictions of the resistance sites indicate progress in the 
development of successful models of real viral evolution at the single residue 
level, and suggest that our approach may be applied to help design new therapies 
that are less prone to failure even where resistance data is not yet available. 
\end{abstract}

\maketitle

Under selective pressure from sub-optimal anti-retroviral treatment regimens, HIV 
has been observed to evolve drug resistance within weeks of treatment initiation 
\citep{MOLL96}. While modern combination therapies have greatly reduced the rate 
of evolution of drug resistance, resistant strains are found in greater than 
14\% of newly infected patients in the United States \citep{VOLB10, WHEE10}. 
The rapid evolution of resistance is congruent with the overall observation that 
HIV evolution is remarkably fast, with studies indicating that in the absence of 
treatment a single patient's HIV infection will explore every possible point 
mutation many times daily \citep{RAMB04, COFF95, PERE96}. However, empirical 
studies of viral sequence data indicate that HIV evolution is structured and 
exhibits reproducible patterns \citep{MOLL96,DAHI11}.  

The existence of significant correlations in the evolution of HIV suggests that 
sequence data can be used to parameterize statistical mechanical models of HIV evolution 
that predict important features of its evolution, including the evolution of drug resistance. 
Previous researchers have used a variety of approaches to predict HIV fitness and aspects 
of its evolution using viral sequence data on its own \citep{DAHI11, FERG13}, and 
with additional phenotypic properties such as drug resistance and replicative 
capacity \citep{HINK11}. Others have addressed the problem of predicting the 
sites of drug resistance mutations by detecting sites under positive selection 
during treatment \citep{CHEN04}, supervised learning \citep{CAO05}, and structural 
modeling of protein-drug interactions \citep{CAO05, BEER02}. 

Here we use HIV sequence data, obtained prior to the widespread clinical use of protease 
inhibitors, to parameterize a spin representation of the standard Eigen 
model of quasi-species evolution \citep{EIGE71, LEUT86, SHEK13}. 
We then use this model to predict sets of sites in HIV protease where joint 
mutations are unlikely to significantly impair viral fitness. We hypothesize that 
such sites are more likely to be sites of clinically relevant drug resistance 
mutations because resistance mutations that severely impair viral replication are 
unlikely to be selected. Our successful identification of major drug resistance 
sites (defined in \citep{RHEE2003}) using natural evolution data suggests that 
our techniques could be applied to predict HIV evolution in response to new 
treatment regimens or vaccine candidates when resistance information is unknown. 

We begin by inferring an estimate of the probability distribution of mutations 
in the viral protease from sequence data. Protease amino acid sequences are 
first translated into a binary form, with the amino acid at each site $i$ encoded by
$s_i\in\{0,1\}$, where 0 (1) denotes a wild type (mutant) amino acid at that site. 
Full sequences are thus represented as vectors $s=(s_1,s_2,..,s_{99})$. We assume
that the joint distribution of mutations is adequately captured by the moments 
$\langle s_i s_j\rangle$ and find the maximum entropy distribution consistent 
with the observed moments (note that because $s_i^2=s_i$, $\langle s_i\rangle = 
\langle s_i^2\rangle$ all first moments are included) \citep{JAYN57, FERG13}. The 
resulting probability distribution takes the form
\begin{align}
\begin{aligned} \label{1}
P(s) &= Z^{-1}\exp(-E(s))\\
E(s) &= \sum_{i<j}^L J_{ij}s_i s_j +\sum_{i=1}^L h_i s_i\,,
\end{aligned}
\end{align}
where $Z$ is the partition function. The parameters $\{J_{ij}\}$, $\{h_i\}$ must 
be chosen such that the distribution $P(s)$ reproduces the observed moments 
$\langle s_i s_j\rangle$. Here the $\{J_{ij}\}$ can be thought of as capturing 
direct interactions between sites, disentangled from the network of correlations 
that include indirect effects mediated through intermediate sites \citep{COCC11, MORC11, SCHN06}. 
Similar maximum entropy approaches have been fruitfully applied to analyze patterns of neural 
activity and to predict contact residues in protein families \citep{SCHN06, WEIG09, MORC11, MARK12}. 
The description of the selective cluster expansion algorithm used to 
infer $E(s)$ is given in \citep{COCC11, BART13}. Although only the pair correlations are 
constrained in Eq.~\ref{1}, the inferred Ising model accurately predicts higher order 
correlations as well. 

\begin{figure}[]
\centering
\includegraphics[width=0.95\columnwidth]{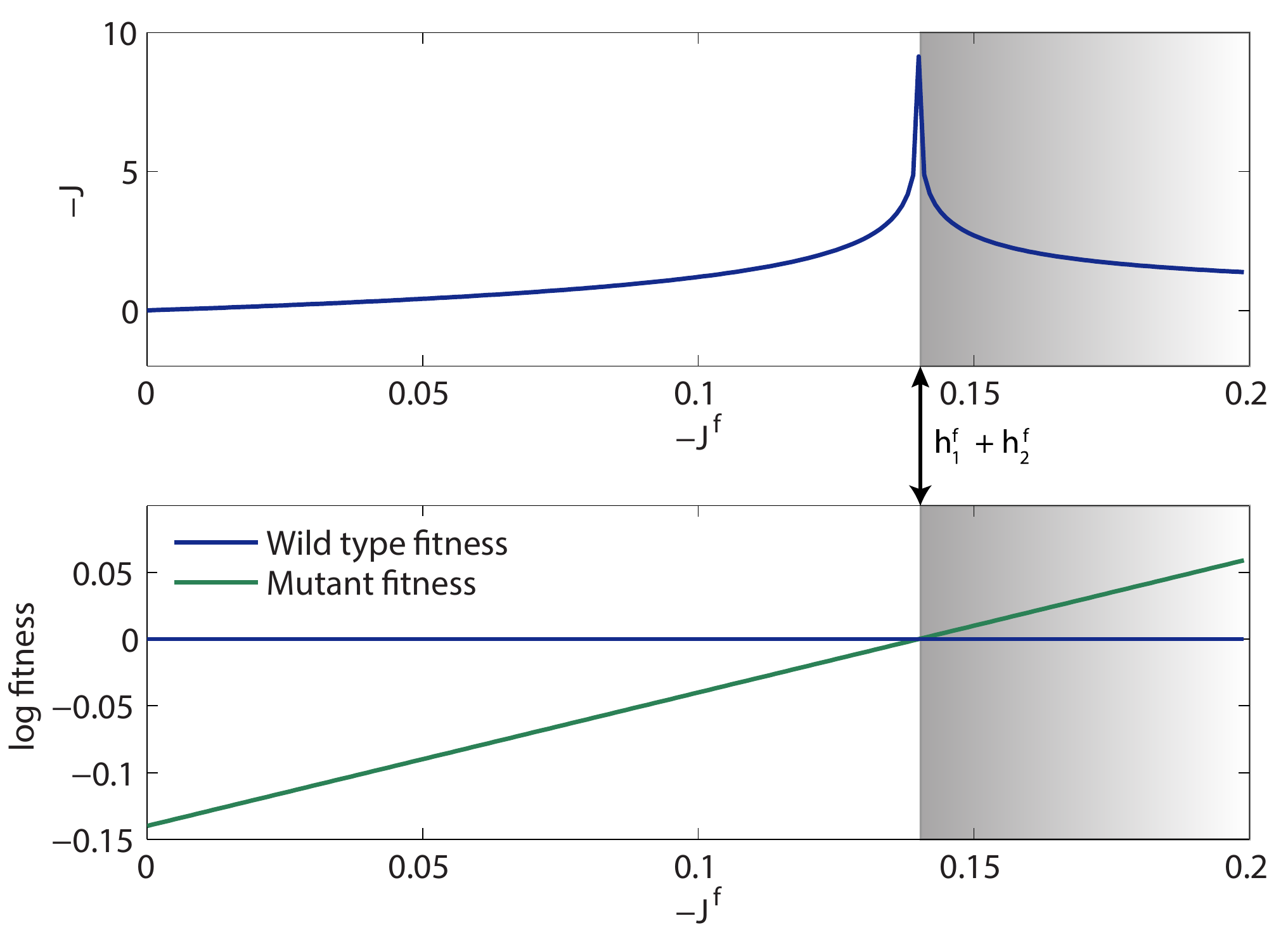}
\caption{The coupling $-J$ between a pair of sites increases sharply as the fitness 
of the double mutant approaches the fitness of the wild type sequence. 
Top panel: The peak in $-J$ occurs at the level crossing, where 
$-J^f = h_1^f+h_2^f$. If $-J^f$ becomes larger 
than $h_1^f+h_2^f$, so that the double mutant has higher fitness than the wild 
type sequence (shaded region), the corresponding coupling $-J$ in the prevalence 
landscape decreases. Bottom panel: Log fitness of the wild type strain and the 
double mutant strain as a function of $J^f$. The log fitnesses intersect at the 
point where $-J$ is maximized.}
\label{fig:crossing}
\end{figure}

The form of the probability distribution gives rise to the notion of a 
``prevalence landscape'' that expresses the relative frequencies of protease 
sequences. Previous work has shown that the inferred prevalences of sequences 
from HIV Gag proteins correlate with their replicative capacities, another proxy 
for fitness \citep{FERG13, MANN14}, in line with the intuition that fitter strains 
should be more prevalent. However, prevalence is affected by many 
factors other than fitness, including epidemiological dynamics, recombination, 
and demographic noise, which complicate this association \citep{GREN04, LEME06, 
NEHE10}.

Insight into the relation between fitness and prevalence can be obtained through 
Eigen's model of evolution \citep{EIGE71}. This model assumes an infinite 
population of viruses, and accounts for mutation and selection, but neglects 
many of the important effects described above. However, these simplifications 
allow for the relationship between fitness and prevalence to be studied 
using methods adopted from 
statistical physics \citep{LEUT86, SHEK13}. Following Eigen's model, the 
prevalence can also be written as the outcome of evolutionary dynamics over a 
large number of generations $T$, represented as a series of coupled Ising spin systems \citep{LEUT86}
\begin{align}
\begin{aligned} \label{2}
\exp(-E(s^T)) \propto \\
\sum_{\{s^t\}_{t=1}^{T-1}}\exp{\left[\sum_{t=1}^{T-1} 
K(2s^t-1)(2s^{t+1}-1)-F(s^t)) \right]} \\
F(s) = \sum_{i<j}^L J^f_{ij}s_i s_j+\sum_{i=1}^L h_i^f s_i,
\end{aligned}
\end{align}
where $K$ is related to the per site per generation mutation rate $\mu$ by $K = 
\frac{1}{2}\log(\frac{1-\mu}{\mu})$. Here $F(s)$ is minus the log fitness of 
sequence $s$ and will be referred to as the fitness landscape. The superscripts 
$t\in\{1,2\ldots,T\}$ on the sequence vectors refer to discrete generations. The 
superscript $f$ on the parameters $\{h^f_i\}$ and $\{J_{ij}^f\}$ indicates that 
these parameters are taken from the fitness landscape in Eq.~\ref{2} (assumed to 
have the same functional form as Eq.~\ref{1}), rather than the prevalence 
landscape of Eq.~\ref{1}. The evolutionary dynamics described here applies to 
evolution within a population of hosts. Equations describing within host 
evolution would require accounting for differing immune pressure between 
individuals \citep{SHEK13}, though protease is not comparatively immunogenic 
\citep{BARTH13}.

Ideally, one would like to invert Eq.~\ref{2} to solve for $F(s)$ 
in terms of $E(s)$, because $E(s)$ is inferred directly from data. 
In principle, this could be achieved by matching the distribution of sequences at the
final generation $T$ in the Ising representation of Eigen's model with the prevalence
landscape, given by Eq.~\ref{1}. This is a challenging problem in general, however, 
approximate results can be obtained by studying a two site system, which can be solved by 
straightforward transfer matrix methods. While network effects influence the inferred couplings between 
sites, this simple approximation provides useful intuition. Furthermore, network effects
exert a weaker influence on the $\langle s_{i} \rangle$, as most of their variance is explained by 
the single site $h_i$ in Eq.~\ref{1}.

Solving the two site version of Eq.~\ref{2} shows that the $h^f$ are difficult to 
reliably infer, because the mutation coupling $K$ is large enough ($K\simeq 
-\log(\mu)$ and $\mu \simeq \mathcal{O}(10^{-4})$) that very small $h^f$ lead to 
large $h$ in the prevalence landscape. However, large values of $J$ in the 
prevalence landscape have a simple interpretation in the fitness landscape as 
couplings between pairs of sites where mutating both sites leads to only a small 
change in fitness compared to wild type  (Fig.~\ref{fig:crossing}). In this case 
the double mutant could become advantageous with only a small increase in the 
fitness of one of the mutations, as might occur when drugs are added to the 
environment, for example. Mathematically, this occurs as $-J^f$ approaches 
$h_1^f+h_2^f$. We refer to the point in parameter space where the coupling 
between sites allows the double mutant strain to have equal fitness to the wild 
type as a level crossing.

To go from the interpretation of large values of $-J$ in the prevalence 
landscape as indicators of nearby level crossings to predictions of resistance 
mutations requires elucidating a relationship between level crossings and 
resistance mutations. A rigorous argument relating resistance mutations to the 
fitness landscape would require detailed knowledge of the drug, its binding 
sites, the structure of the target protein, and other details. 
However, generically we expect that when the environment in which HIV  
replicates changes due to the initiation of drug therapy, HIV must mutate in 
ways that abrogate drug binding, while at the 
same time preserving protein function. Large couplings $-J$ connect sites 
that are likely to be able to co-mutate with limited costs to fitness, even if 
the associated individual mutations are costly. Such sets of sites are therefore 
more likely to be associated with resistance. Here our assumption is that 
resistance cannot be achieved through selectively neutral mutations at single 
sites, in which case drug treatment would likely be ineffective. 

\begin{figure}[]
\centering
\includegraphics[width=1.00\columnwidth]{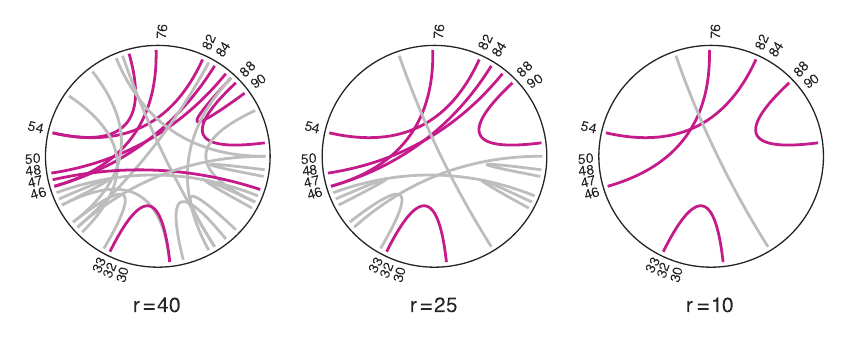}
\caption{Stronger couplings are more likely to link sites of major resistance 
mutations. Here we show the network of interactions between the top $r$ ranked 
sites, from $r=40$ (left) to $r=10$ (right). Only the strongest couplings, 
those meeting or exceeding the largest coupling for the lowest ranked site, are 
displayed. Interactions linking at least one major resistance site are darkly 
shaded, links between non-resistance sites are lightly shaded.}
\label{fig:network}
\end{figure}

To predict the sites of resistance mutations based on the above considerations, 
we consider the strongest couplings $-J_{ij}$ associated with each site $i$. 
Using the largest coupling values we then assign each site a rank 
$r\in\{1,\ldots,99\}$ from strongest to weakest. We predict that the sites with 
the strongest interactions (i.e.~the highest ranked sites) are most likely to be 
associated with drug resistance. Focusing on the highest ranked sites, and the 
strong couplings between them, can be seen as a process of pruning weaker 
interactions from the network. Three pruned versions of the network of 
mutational interactions in HIV protease are shown in Fig.~\ref{fig:network}.

\begin{figure}[]
\centering
\includegraphics[width=0.95\columnwidth]{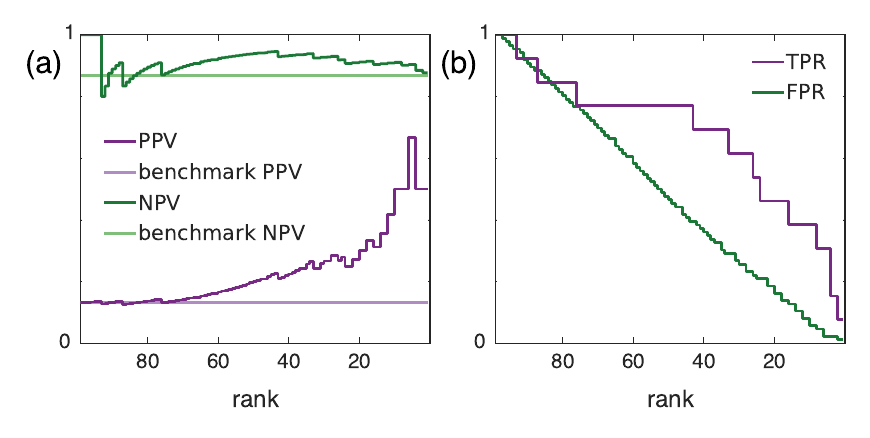}
\caption{Top-ranked sites, based on the maximum strength of their couplings, are 
far more likely to be sites of major drug resistance mutations than would be 
expected by chance. (a) Positive prediction value (PPV) and negative 
prediction value (NPV) for the classifier compared to the benchmark of random 
guessing, as a function of rank. Collections of the highest ranked sites are 
clearly associated with improved PPV. (b) False positive rate (FPR) 
and true positive rate (TPR) as functions of rank. TPR $>$ FPR indicates 
performance better than chance.}
\label{fig:classifier}
\end{figure}

This model can be cast in the form of a classification rule by predicting 
sites ranked at or above some threshold rank $r$ to be sites of drug resistance 
mutations, and sites of lower rank to be unassociated with resistance. To test 
the model's performance, we take the set of resistance sites to be those 
classified as sites of major resistance mutations by the Stanford HIV drug 
resistance database (sites 30, 32, 33, 46, 47, 48, 50, 54, 76, 82, 84, 88, and 
90) \citep{RHEE2003}. As higher ranked sites are selected, the proportion of sites 
that are associated with resistance should increase. This can be measured using 
positive prediction value (PPV) and negative prediction value (NPV), defined as 
\begin{align}
\begin{aligned}
&P(\text{true = resistance}|\text{predicted = resistance})\,, \\ \nonumber
&P(\text{true = non-resistance}|\text{predicted = non-resistance})\,. \nonumber
\end{aligned}
\end{align} 
These are shown in Fig.~\ref{fig:classifier} compared to benchmarks 
for a random classifier, and demonstrate that the performance of the 
classification rule is substantially better than chance for higher ranked sites. 
Examination of the true positive rate (TPR) and false positive rate (FPR), 
\begin{align}
\begin{aligned}
&P(\text{predicted = resistance}|\text{true = resistance})\,, \\ \nonumber
&P(\text{predicted = resistance}|\text{true = non-resistance})\,, \nonumber
\end{aligned}
\end{align}
shown in Fig.~\ref{fig:classifier}, confirm that TPR$>$FPR, indicating 
performance better than chance. We also note that the fraction of the strongest 
interactions which link at least one major drug resistance site is extremely 
high, as can be seen in Fig.~\ref{fig:network} (further details in Supplemental 
Material).

To examine these results using classical statistical significance testing, we 
used the hyper-geometric distribution to compute p-values for the null model of 
randomly selecting the number of sites at or above each rank threshold and 
obtaining at least as many resistance mutations as found using the ranking 
classifier. The predictions have p-values $<0.05$ for essentially all rank 
thresholds from $r=3-50$, which comports with the argument that strongly coupled 
sites are more likely to be sites of resistance mutations and supports the 
significance of the predictions of resistance among higher ranked sites. The 
lack of significance for the highest ranked pair is a consequence of the very 
small number of sites. We also tested related classification rules constructed 
using direct information \citep{MORC11} and correlation matrices, with no 
improvement in performance (Supplemental Material).

As virological failure occurs in patients undergoing treatment with protease 
inhibitors, new protease inhibitor drugs are administered \citep{VOLB10}. To 
further assess the validity of our predictions, we used the model to infer pairs 
of protease inhibitors that are optimized to protect patients from evolving 
multi-drug resistance. To protect against resistance, a pair of drugs should 
have as many non-overlapping resistance mutations as possible. Additionally, the 
drugs' associated resistance mutations should be difficult to make simultaneously 
due to fitness constraints. In the same way 
that large positive values of $-J$ indicate sites that can readily mutate 
together, negative values of $-J$ indicate sites where double mutations are 
suppressed. Thus, the interactions between the resistance mutations that are not 
common to both drugs should be as negative as possible. We found three 
combinations (atazanivir-indinavir, atazanavir-fosamprenavir, and 
darunavir-nelfanavir) that are optimal for both of these criteria in the Pareto 
sense: improvement in one criterion necessitates a reduction in the other 
criterion. Two of these, along with both near-optimal pairs 
(atazanavir-darunavir and atazanavir-lopinavir), incorporate atazanavir, 
consistent with clinical knowledge that the resistance profile of atazanavir 
appears distinct from other protease inhibitors \citep{COLO04}.

The network of large interactions also captures important 
biophysical information. As a first example, the third strongest coupling is 
between sites 82 and 54. Site 82 is frequently the first resistance mutation 
site observed after the initiation of protease inhibitor treatment, and is 
usually followed by mutation at site 54 \citep{MOLL96}. Some couplings may also 
be associated with stabilizing mutations, which compensate for loss of fitness 
due to a destabilizing mutation. A recent biophysical study examined the melting 
temperatures of HIV protease with a major resistance mutation at site 84 
\citep{CHAN11}. The study showed that on its own, the major resistance mutation 
reduced the stability of HIV protease considerably. When the mutation at site 84 
is accompanied by one of a set of three known accessory mutations at sites 10, 
63, and 71, stability is restored, or even enhanced. Couplings between sites 10 
and 84, and sites 63 and 84, are strong, in the top 7\% of all couplings (though 
weaker than the couplings shown in Fig.~\ref{fig:network}, which are within the 
top 1\%). The coupling between sites 71 and 84 is slightly weaker, but still in 
the top 13\% of all couplings. This suggests that links between destabilizing 
mutations and those that improve protein stability may be captured by the 
network of interactions inferred from sequence data.

Our results show that from sequence information alone, much of the evolutionary 
response of HIV protease to inhibitors can be reproduced. While in the case of 
protease inhibitors, the answer was known, the successful retrodictions indicate 
that our understanding of HIV evolution is becoming predictive at the 
level of individual residue sites. We anticipate that the methods developed 
above will contribute to the development of predictive theories of viral evolution and
to the development of new treatments, such as integrase 
inhibitors \citep{POMM05}, where resistance is not nearly as well characterized 
as in protease.

We thank Daniel Kuritzkes, Martin Hirsch, Andrew Ferguson, Dariusz Murakowski, and Hanrong Chen 
for helpful discussions. This research was funded by the Ragon Institute of MGH, 
MIT, \& Harvard, and NSF under Grants No.~PHY11-25915 and DMR-12-06323.

\bibliography{HIV_protease}

\end{document}